\documentclass[paper]{JHEP3} 

\usepackage{epsfig,multicol,bbm}

\newcommand\fverb{\setbox\fverbbox=\hbox\bgroup\verb}
\newcommand\fverbdo{\egroup\medskip\noindent%
			\fbox{\unhbox\fverbbox}\ }
\newcommand\fverbit{\egroup\item[\fbox{\unhbox\fverbbox}]}
\newbox\fverbbox

\newcommand{\mpl}{M_{\rm Pl}}
\newcommand{\veff}{V_{\rm eff}}
\newcommand{\feff}{f_{\rm eff}}
\newcommand{\eff}{{\rm eff}}

\title{Orbifold GUT inflation}

\author{Seong Chan Park\\
	FPRD, School of Physics and Astronomy, Seoul National University,\\
    Seoul 151-747, Korea\\
	E-mail: \email{spark@phya.snu.ac.kr}}

\abstract{We consider a scenario of cosmological inflation coming from a grand unified theory
in higher dimensional orbifold. Flatness of the potential is automatically guaranteed in this orbifold setup
thanks to the nonlocality of the Wilson line on higher dimensions and the local quantum gravitational corrections
are exponentially suppressed. The spectral index of scalar perturbation ($n_s \simeq 0.92 - 0.97$) and
a significant production of gravitational waves are predicted ($r= T/S \simeq 0.01 - 0.12$) in the perturbative regime
of gauge interaction, ($1/g_4 = (5 \thicksim 20)\times 2 \pi R \mpl$) where the size of compactification
is constrained ($ R \mpl \simeq 20- 45$) by the measurement of scalar power spectrum ($\Delta_{\cal R} \simeq 5 \times 10^{-5} $).}


\begin{document}

\section{Introduction}
Inflation is the best known idea solving several cosmological puzzles of the standard big bang cosmology such as
homogeneity, isotropy, and flatness of the universe \cite{Guth:1980zm, Linde:1981mu, Albrecht:1982wi}.
Recent precise observations of cosmic microwave background radiation (CMBR) strongly support the actual existence of the
epoch of accelerated expansion in early universe \cite{Spergel:2006hy}.

In particle physics models, inflation is driven by a single (or multiple) scalar field(s) dubbed inflaton field(s)
and the potential for the field is required to be nearly flat to provide enough time for exponential expansion or e-foldings. To protect the flat potential from quantum corrections, models often constructed in the context of symmetry principle such as supersymmetry (for a review, see \cite{Lyth:1998xn}) and an axionic shift symmetry (for a recent review, see \cite{Savage:2006tr}). However it seems that none of suggested
particle physics models are entirely convincing because of many reasons. In supersymmetric models, generic supergravity correction can spoil the flatness of the inflaton potential during the inflation since it can induce a large mass correction for inflaton field \cite{Copeland:1994vg}. Other models of inflation associated with the shift symmetry also are not fully satisfactory since they often require the trans-Planckian fluctuations of inflaton field ($\delta \phi \gtrsim \mpl = \sqrt{8 \pi G}= 2.4 \times 10^{18}{\rm GeV}$) \cite{Freese:1990rb,Adams:1992bn}. A sole symmetry principle may not provide a fully acceptable framework for inflation and there might be something beyond it \cite{Kim:2004rp}.

A recent attempt made by Arkani-Hamed et.al. \cite{Arkani-Hamed:2003wu} (see also \cite{Arkani-Hamed:2003mz}) based on higher dimensional spacetime could be interesting because the inflaton potential in their consideration is automatically free from quantum gravitational corrections thanks to the nonlocality of higher dimensional construction itself. In their original construction, they consider a five dimensional $U(1)$ gauge theory on $M^4 \times S^1$ where $M^4$ denotes the Lorentz spacetime and $S^1$ a circle compactification with a length $L$. First, they consider a nonlocal operator defined by the gauge invariant Wilson line,
\begin{eqnarray}
e^{i \theta} = e^{i \oint A_5 dy}.
\end{eqnarray}
Below the $1/L$ scale, the dynamics of the Wilson line field $\theta$ is described by a Lagrangian
\begin{eqnarray}
{\cal L} = \frac{1}{2 g_4^2 L^2 } (\partial \theta)^2 - V(\theta) + \cdots,
\end{eqnarray}
where $g_4^2 = g_5^2/L$ is the effective four dimensional coupling constant.
At one loop level, the potential $V(\theta)$ is induced by interactions with the charged bulk fields (of charge $q$) as
\begin{eqnarray}
V(\theta) \simeq \pm \frac{{\rm Const.}}{L^4} \sum_{n=1}^\infty \frac{\cos{(n q \theta) }}{n^5}
\end{eqnarray}
where the sign depends on spin of interacting particle. The effective potential essentially has the same form with Natural inflation \cite{Freese:1990rb,Adams:1992bn} with the effective decay constant given by $\feff= 1/ (g_4 L)$. One should note that
no dangerous higher-dimensional operator can be generated in a local higher dimensional theory and the potential can be
trusted even when $\feff$ can be larger than $\mpl$ in the perturbative regime of gauge interaction
\footnote{We noticed a ref. \cite{Arkani-Hamed:2006dz} where authors claimed that $U(1)$ gauge (without light charged particles)
could not be compatible with string theory up to arbitrarily high energy and there should appear a cutoff scale around
($\Lambda \sim g \mpl $) even the theory seems perfectly fine as a low energy effective theory.}. Of course, this original
model with the $U(1)$ symmetry should be considered as a toy model and we have to construct a realistic model where the standard model gauge group $G_{\rm SM}=SU(3)\times SU(2) \times U(1)$ is fully considered as an effective field theory. That's the main motivation of the current work.

In this paper, we try to get a potentially realistic model of inflation from a higher dimensional gauge theory.
Having the standard model as a low energy effective theory, we would consider the $SU(5)$ gauge theory as the 
starting point. Orbifold projection can provide a nice explanation of doublet-triplet splitting \cite{Kawamura:2000ev}.
We found that only one particular choice of orbifold projection is available to get the standard model as well as 
the inflaton field. This will be clarified in the following sections. 
We would emphasize one more nice property of this orbifold construction: Distinguishing from the $U(1)$ toy model,
the non-Abelian nature of $SU(5)$ GUT allows the nontrivial one-loop potential of inflaton field solely coming from the
gauge self interactions. Without introducing any (arbitrary) charged bulk field in the model, no further ambiguity occurs in predicting cosmological observable quantities. Spectral index ($n_s = 0.92 \thicksim 0.97$) and
a significant production of gravitational waves is predicted ($r= T/S = 0.02 \thicksim 0.12$). No large non-Gaussianity
is expected by the model.

The content of this paper is given as follows. Next section, we specify the $SU(5)$ GUT model based on $S^1/\mathbb{Z}_2$ orbifold compactification. The orbifold boundary condition is chosen in such a way that the standard gauge bosons and a massless scalar particle
remain after the orbifold compactification. The massless scalar particle, coming from the fifth component of the gauge boson ($A_5$), defines 
a Wilson-line phase and its radiative potential is calculated at one-loop level. 
In Section 3, we analyze the potential and show that the slow-roll conditions are nicely satisfied when the gauge coupling is weak. 
Enough time for efoldings is obtained. Various cosmological observable quantities such as the spectral index of the scalar perturbations, $n_s$, the ratio of tensor to scalar perturbations, $r=T/S$, and the running of the spectral index, $d n_s/d\ln k$, are fully predicted. In the last section, a summary is given.


\section{The model}
In this section, we specify the model of cosmological inflation coming from the $SU(5)$ GUT gauge theory on
the  orbifold ($M^4 \times S^1/\mathbb{Z}_2$). Let us assume that only pure gauge fields (and graviton) are propagating
through the bulk and all the fermion fields are localized on one of the fixed points of $S^1/\mathbb{Z}_2$
orbifold \footnote{Actually one-loop generated effective potential for the inflaton, $A_5$, could be induced
solely by the gauge self-interaction without any further requirement of charged bulk fields. For the simplicity
and predictability, here we assumed that all the fermions are localized and do not directly affect the one-loop
effective potential at the leading order.}. The coordinates of five dimensional space-time
are denoted by $x^M = (x^\mu, y)$ where the indices are given as
($M=0,1,2,3,4, \mu=0,1,2,3$), respectively. We require that the Lagrangian is single valued and gauge invariant.
To respect the orbifold $\mathbb{Z}_2$ condition, boundary conditions for bulk fields should be specified
by two parity matrices, $P_0$ and $P_1$ around fixed points, $y=0$ and $y=\pi R$, respectively \footnote{Generically, one could
specify three conditions to get an invariant theory on $S^1/\mathbb{Z}_2$ orbifold. One for translation, $U: y \rightarrow y+ 2\pi R$,
two for $\mathbb{Z}_2$-orbifold conditions around fixed points, $P_0: y \rightarrow -y$ and $P_1: y+\pi R \rightarrow -y+\pi R$, respectively.
However, as a transformation $y+\pi R\rightarrow -y+\pi R$ must be the exactly same as a transformation $ y + \pi R \rightarrow -(y+\pi R)
\rightarrow -y +\pi R$, it follows that $U = P_1 P_0$ or the translation can be obtained by two parity operations.}.
\begin{eqnarray}
P_0:&& A_M (x^\mu, -y) = (-1)^\alpha P_0 A_M (x^\mu, y) P_0^\dag, \nonumber \\
P_1:&& A_M (x^\mu, \pi R -y) = (-1)^\alpha P_1 A_M (x^\mu, \pi R + y) P_1^\dag
\label{conditions}
\end{eqnarray}
where $\alpha =0 (1)$ for $M =\mu (5)$, respectively. The extra $(-1)$ sign is required to preserve the Lorentz invariance.
Taking care of overall sign ambiguities, there are essentially two independent ways to break the GUT gauge group,
$SU(5)$, down to the standard model gauge group, $SU(3)\times SU(2) \times U(1)$ \cite{Haba:2003ux, Haba:2004qf}.
\begin{itemize}
  \item Choice-I: %
\begin{eqnarray}
P_0 =\left(
        \begin{array}{cc}
          -I_3 & 0 \\
          0 & I_2 \\
        \end{array}
      \right),
P_1 =\left(
        \begin{array}{cc}
          I_3 & 0 \\
          0 & I_2 \\
        \end{array}
      \right),
\end{eqnarray}
  \item Choice-II:
  \begin{eqnarray}
P_0 =\left(
        \begin{array}{cc}
          -I_3 & 0 \\
          0 & I_2 \\
        \end{array}
      \right),
P_1 =\left(
        \begin{array}{cc}
          -I_3 & 0 \\
          0 & I_2 \\
        \end{array}
      \right).
\end{eqnarray}
\end{itemize}
Here $I_n$ denotes $n\times n$ unit matrix.
Applying the parity matrices to Eqs. \ref{conditions}, we could read out the parity assignment for the $SU(5)$ gauge
boson in adjoint $5\times 5$ matrix representation as follows.
\begin{itemize}
  \item Choice-I:
\begin{eqnarray}
A_\mu = \left(
          \begin{array}{cc}
            (++) & (+-)\\
            (-+) & (++) \\
          \end{array}
        \right),
A_5 =\left(
       \begin{array}{cc}
         (--) & (-+) \\
         (+-) & (--) \\
       \end{array}
     \right),
\end{eqnarray}
  \item Choice-II;
  \begin{eqnarray}
A_\mu = \left(
          \begin{array}{cc}
            (++) & (--)\\
            (--) & (++) \\
          \end{array}
        \right),
A_5 =\left(
       \begin{array}{cc}
         (--) & (++) \\
         (++) & (--) \\
       \end{array}
     \right).
\end{eqnarray}
\end{itemize}

In either choices, there appear massless
gauge bosons, with $(++)$ parity, having exactly same quantum numbers with the standard model gauge bosons, i.e., gluons, $W^\pm$,
$W^3$ and $B$. The main difference between the first and the second choice is the existence of
massless scalar degree of freedom. Only the second choice allows the massless scalar degree of freedom
with $(++)$ parity, in $A_5$. Indeed, Kawamura \cite{Kawamura:2000ev,Kawamura:2000ir} took
the first choice, to address the doublet-triplet splitting problem of GUT and he could recover the standard
model without any light exotic scalar degree of freedom at the lowest level of Kaluza-Klein decomposition. In the first choice the lightest scalar field acquires non-zero mass ($\sim 1/R$) since it has $(+-)$ or $(-+)$ parity instead of $(++)$ parity. However, we {\it need} a light scalar field for inflation as well as the standard model gauge fields. Thus our choice is the second one!

This would-be-inflaton scalar field can be written in canonical normalization as follows:
\begin{eqnarray}
\Phi = A_5 \sqrt{\pi R} = f \left(
                                                \begin{array}{cc}
                                                  0 & \phi \\
                                                  \phi^\dag & 0 \\
                                                \end{array}
                                              \right).
\end{eqnarray}
Here  we introduced a mass scale parameter $f \equiv 1/(2\pi R g_4)$ in such a way that the field $\Phi$ 
is canonically normalized with a dimensionless angle parameter $\phi$ where the four dimensional effective coupling is  $g_4^2 = g_5^2/\pi R$.
One should note that the scale, $f$, can be {\it large} at the weak coupling limit, $g_4 \ll 1$. This feature is indeed
a unique feature in higher dimensional gauge theory contrast to the case in four dimensional theory where effective
energy scales should be considered less than Planck scale.

Now let us define a Wilson line phase, $W = P e^{i \int_C A_5 dy}$.
The Wilson line can be parameterized by two independent real numbers after taking account of the remaining symmetry after orbifold
projection.
\begin{eqnarray}
\phi = \left(
         \begin{array}{cc}
           \alpha  & 0 \\
           0 & \beta  \\
           0 & 0 \\
         \end{array}
       \right).
\end{eqnarray}
The effective potential for $\Phi$ can be evaluated (see \cite{Haba:2004qh} for the general formula of
the effective potential in 5D $SU(N)$ gauge theory) provided that the full particle spectra of the theory is
specified. Here we just turn on the bulk gauge sector without further complication potentially
coming from the fermionic sector. After taking care of massless gauge fields and ghost fields,
the total effective potential is obtained as
\begin{eqnarray}
V_\eff (\Phi) = \frac{1}{R^4}\left( c(\alpha) + c(\beta) + \frac{c(2\alpha)+c(2\beta)}{2}+c(\alpha+\beta)+c(\alpha-\beta)\right)
\end{eqnarray}
dropping the divergent cosmological constant term \footnote{In this sense, we do not address the cosmological constant problem.
We would simply put the potential vanishing at the origin, $V_\eff (0) =0$.}.
A dimensionless one-loop function, c(x), is calculated as follows.
\begin{eqnarray}
c(x) = -\frac{9}{64\pi^6} \sum_{n=1}^\infty \frac{\cos n \pi x}{n^5}.
\end{eqnarray}
The function is periodic, $ c(x) = c(x+ 2\pi)$, and even, $c(-x) = c(x)$ with respect to inversion, $x\rightarrow -x$,
along the compact fifth direction. Let us check the effective potential more closely (see Fig.1). First, we can notice that the global minimum of the potential locates at the origin, $(\alpha,\beta)=(0,0)$ and the gauge symmetry, $SU(3)\times SU(2)\times U(1)$ is intact even after taking care of one-loop corrections. If the initial point of the inflaton fluctuation belongs to local maxima or below the maxima, the field is going slowly down
to the global minimum where the standard gauge symmetry is fully recovered. This is an interesting observation since the setup does not require
any fine-tuning in the parameter space or even any arbitrarily extended sector is required beyond the assumption of the minimal $SU(5)$ GUT with a properly chosen boundary conditions.

\begin{figure}[ht]
\begin{center}
\includegraphics[width=.65\linewidth]{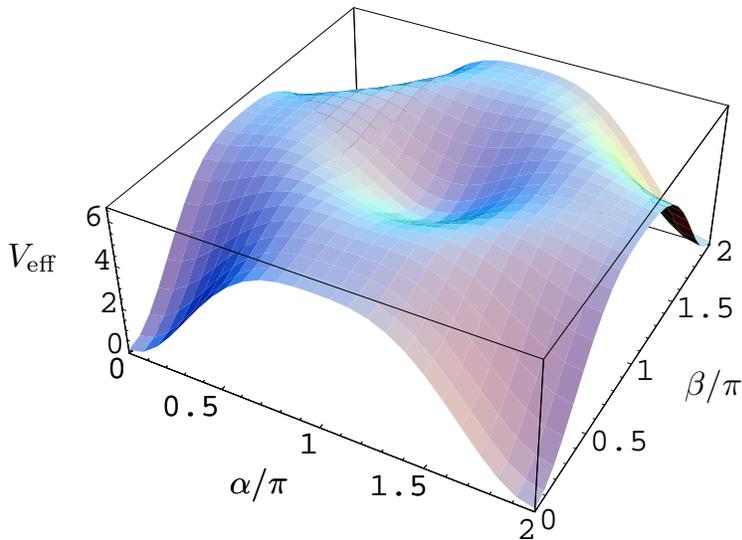}
\end{center}\caption{The one-loop effective potential of $\Phi$ in the unit of ($9/64\pi^6R^4$). The inflaton field is going down to the origin
where the standard gauge symmetry intact. The potential is typically flat enough once the coupling constant is weak enough.}
\end{figure}

The cosmological scenario is now obtained as follows. In the beginning, the universe started from a (false) vacuum where the space is a five dimensional orbifold. A gauge theory of $SU(5)$ GUT dictated the fundamental symmetry of particle interactions and also provided a scalar field as a
part of $A_5$ component which could play a role as the slowly-rolling inflaton field. Below the compactification scale, $1/R$, heavy modes beyond the standard model were decoupled. The inflaton field slowly ran down to the global minimum where the standard model gauge symmetry
was fully recovered \footnote{During the inflation, it is expected that some amount of Baryon number could be produced
since the VEV of potential is not vanishing. However its density could hardly affect the current Baryon asymmetry because of dilution. We thank J.E.Kim for indicating this point.}. After inflation, usual Higgs phase opened and the standard electroweak symmetry breaking took place with the standard
Maxican Hat potential for the Higgs field (see ,e.g., \cite{Cacciapaglia:2005da} for electroweak symmetry breaking from an orbifold gauge theory).


\section{Efoldings, Slow roll parameters and predictions}

Given inflaton potential, we now predict cosmological observable quantities. Let us first check whether the potential
accommodates the enough time for $60$ efoldings. To check this, we would analyze the potential along
the steepest line along which $\alpha = \beta$. All the other line of rolling path consumes longer time and larger number
of efolding is automatically guaranteed.


\begin{figure}[ht]
\begin{center}
\includegraphics[width=.65\linewidth]{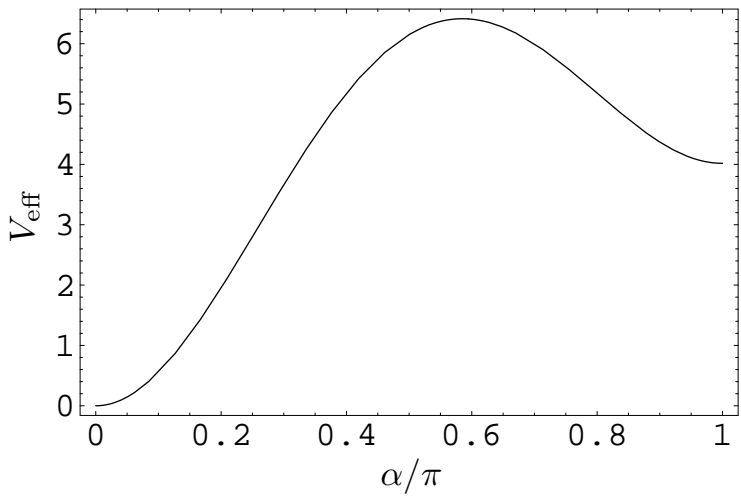}
\end{center}\caption{The one-loop effective potential of $\Phi$ along the steepest line, $\alpha =\beta$.The potential is
slow rolling in the weak coupling limit, $g_4 \ll 1/(2\pi R\mpl)$.}
\end{figure}

The number of efolds, $N_e$ is obtained
by a time integration of Hubble parameter ($H$) and in the slow roll paradigm it can be nicely approximated as follows:
\begin{eqnarray}
N_e = \int_t^{t_{\rm end}} H dt \simeq \frac{1}{\mpl^2}\int_{\Phi_{\rm end}}^\Phi \frac{V_\eff}{V_\eff'} d\Phi
\end{eqnarray}
where $\Phi_{\rm end}$ defines the end of inflation after $60$-efoldings. Here we notice that the number of e-folds is
proportional to the square ratio of the fictitious scale, $f$, and the Planck scale: 
\begin{eqnarray}
N_e \propto \left(\frac{f}{\mpl}\right)^2.
\end{eqnarray}
Again, one should notice that this ratio can
be much larger than $1$ if the coupling is very small, $g_4 \ll 1/(2\pi R \mpl)$.

\begin{figure}[ht]
\begin{center}
\includegraphics[width=.65\linewidth]{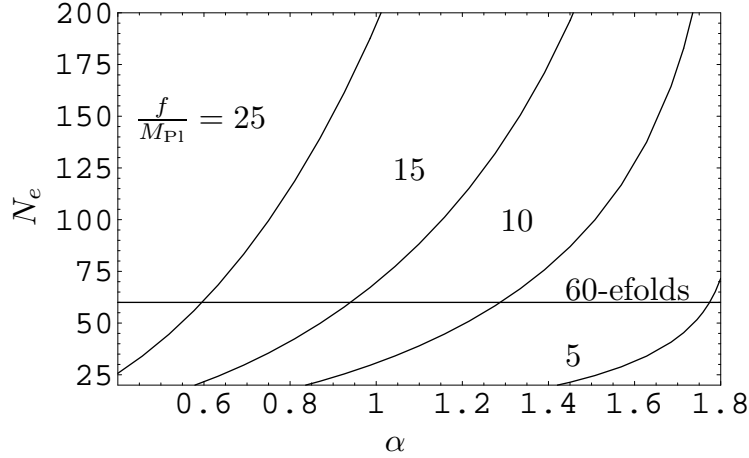}
\end{center}\caption{The number of e-folds, $N_e$, along $\phi \propto (\alpha,\alpha)$  with
respect to the initial value of $\alpha$ for the values $f/\mpl = 20, 25, 30, 35$, respectively. 
For the larger value of $f/\mpl$, it is easier to get the larger number of
e-folds because $N_e \sim (f/\mpl)^2$.}
\label{nefold}\end{figure}

In Fig.\ref{nefold}, we draw a plot for the number of e-folds as a function of the initial value for $a$ where the inflation started. 
Because the number of e-folds is proportional to $(f/\mpl)^2$, we get the larger number of e-folds with the larger value 
of ($f/\mpl$), as is expected. When the inflation started from the top of the potential, $\Phi_{\rm ini}=\Phi_{\rm top}-\Delta_{\rm Quantum} \Phi$ 
with the small quantum fluctuation, we could get sixty efoldings when $f/\mpl \gtrsim 3$.

Let us now check the consistency of slow-roll paradigm and try to make predictions for cosmological observables. In the slow-roll paradigm, three slow-roll parameters, $\epsilon, \eta$ and $\xi$, determine all the observable quantities such as the spectral index, tensor to scalar perturbation ratio, running of spectral index. Slow-roll parameters are defined as follows.
\begin{eqnarray}
\epsilon &=& \frac{\mpl^2}{2}\left(\frac{\veff'}{\veff}\right)^2, \\
\eta &=& \mpl^2 \left(\frac{\veff''}{\veff}\right), \\
\xi &=& \mpl^4 \frac{\veff' \veff'''}{\veff}
\end{eqnarray}
where primes denote derivatives with respect to the field,$\Phi$. Since the ratio, $f/\mpl$, could become
very large as was discussed earlier,  slow roll conditions are easily satisfied:
\begin{eqnarray}
\epsilon, \eta, \sqrt{\xi} \sim \left(\frac{f}{\mpl}\right)^{-2} \ll 1.
\end{eqnarray}
%
\begin{figure}[ht]
\begin{center}
\includegraphics[width=.65\linewidth]{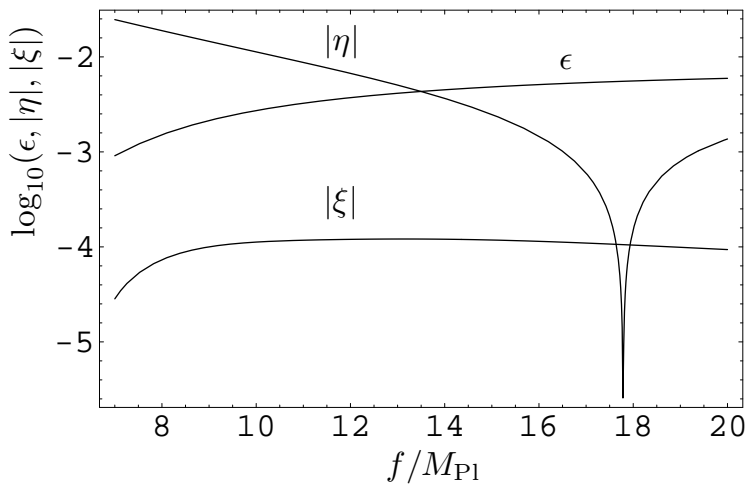}
\end{center}\caption{Slow roll parameters in the orbifold GUT. From the top to the bottom, we draw $\epsilon$, $|\eta|$ and $\xi$
in terms of $f/\mpl$. All of them are small enough and we can consistently rely on slow-roll paradigm for inflation.}
\label{slow roll} \end{figure}

In Fig. \ref{slow roll}, we draw plots for slow-roll parameters, $\epsilon, |\eta|$ and $|\xi|$ with respect to the ratio $f/\mpl$
in the range of $10 \sim 100$ in log scale. It is found that all the slow-roll parameters are sufficiently small
in large range of parameter space and we can safely stay in the slow-roll paradigm.

\begin{figure}[ht]
\begin{center}
\includegraphics[width=.65\linewidth]{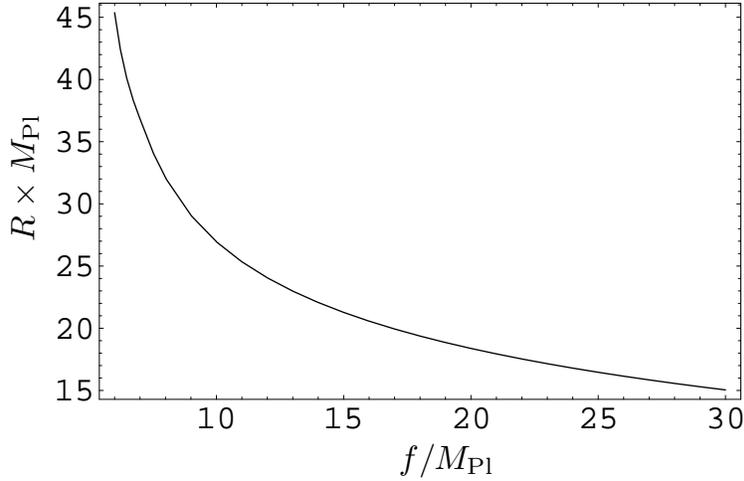}
\end{center}\caption{From the measurement of scalar power spectrum, $\Delta_{\cal R}(k)^2 \approx 5\times 10^{-5}$, we could find
a strict contraint on the size of extra dimension,$R$. The compactification radius is about $15-45$ times larger than the Planck length,$1/M_{\rm Pl}$
in the chosen parameter space for $5 \leq f/M_{\rm Pl} \leq 30$.}
\label{constraint R}
\end{figure}

Cosmological measurements provide an important information about the structure of the inflaton potential. In particular, observational
constraints on the amplitude of scalar perturbations, in the slow roll framework, imply that
\begin{eqnarray}
\Delta_{\cal R}\approx \frac{1}{2\sqrt{3}\pi M_{\rm Pl}}\sqrt{\frac{\veff}{\epsilon}} \approx 5 \times 10^{-5}.
\end{eqnarray}
From this constraint, we could read out the size of extra dimension. In Fig.\ref{constraint R}, the size of extra dimension,$R$, is
calculated in terms of $f/M_{\rm Pl}$.

A standard slow-roll analysis also gives observable
quantities such as the spectral index, $n_s$, the relative contribution of gravitation to scalar perturbation, $r=T/S$
and the running parameter of the spectral index, $dn_s/d \ln k$ in terms of the slow-roll parameters to the first order as
\begin{eqnarray}
n_s &=& 1- 6\epsilon + 2\eta, \\
r &=& T/S = 16 \epsilon, \\
\frac{d n_s}{d \ln k} &=& -16 \epsilon \eta + 24 \epsilon^2 + 2\xi ^2.
\end{eqnarray}

Since $ dn_s/d\ln k$ has double suppression by factor of $ (f/\mpl)^{-4} \ll 1$, it is negligibly small. It is challenging
to measure this small amount of parameter within the range of near future sensitivity.

\begin{figure}[ht]
\begin{center}
\includegraphics[width=0.75\linewidth]{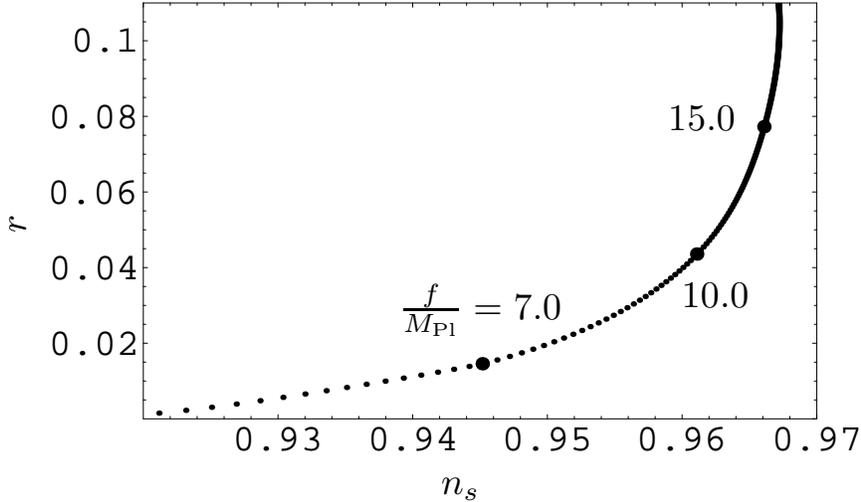}
\end{center}\caption{The orbifold GUT inflation predicts the spectral index of scalar perturbation,$n_s =0.92-0.97$ tensor to scalar ratio, $r=0.01-0.12$ with the assumed parameter space $(5\lesssim f/\mpl \lesssim 20)$.}
\label{prediction}
\end{figure}

In Fig. \ref{prediction}, we plot the prediction for the spectral index ($n_s$) and tensor to scalar contribution($r$) 
for various values of $(f/\mpl)$. With the larger value of ($f/\mpl$), we get the larger spectral index ($n_s \lesssim 0.97$)
and larger tensor-to-scalar ratio $(r \lesssim 0.12)$. Around the point $f/\mpl \sim 15.0$, the spectral index is almost saturated 
with the value of $0.97$ but the tensor-to-scalar ratio seems to go larger and larger.

Here the spectral index, tensor contribution and the running index are well consistent with the recent WMAP data. Interestingly,
the tensor to scalar contribution, $r$, is rather higher than usual KKLT type string theory models where $r$ is negligibly small
when the gravitino mass is given in TeV range \cite{Kallosh:2007wm}. In that sense, the detection of gravitational contribution will be a nice
probe of the orbifold GUT inflation model.

\section{Summary}

In this paper, we present a model of inflation from the $SU(5)$ orbifold GUT model. The inflation field
arises as a consequence of the symmetry transition from a grand unified symmetry, $SU(5)$ to the standard model gauge symmetry,
$SU(3)\times SU(2) \times U(1)$ by the orbifold compactification. Advantage of this model is that the inflaton field is a built-in ingredient 
of the theory and it is  automatically free from local quantum gravitational effects because of its higher dimensional locality and gauge symmetry. 
Fully radiatively induced inflaton potential is naturally slow-rolling
once the theory is weakly coupled during the inflationary era. The spectral index $n_s$ is predicted to be in the range ($0.92 \sim 0.97$) which is fully consistent with the recent observational data.
An interesting prediction is that the significant gravitational wave is expected as $ r \simeq 0.02 \sim 0.12$ when the model parameter is 
assumed in the range of $f/\mpl = 5-20$. Very small running parameter of spectral index is expected as well, $(d n_s/d \ln k \lesssim 0.002)$.
We would leave a study to see the coupling unification in the orbifold GUT models in association with the inflationary scenario proposed in the current paper as a future work.

\acknowledgments{I would thank Andrew Cohen and Tony Gherghetta for useful comments on the validity of
effective theory description. Also we thank K.S. Babu and Sudhir Vempati for encouragement for publication
of the idea. Prof. J.E. Kim and Prof. K. Okada provided an argument on possible production of baryon number during the inflation period.}

\end{document}